\documentclass[pre,twocolumn]{revtex4}
\usepackage{graphicx}
\usepackage{amsmath}
\usepackage{latexsym}
\usepackage{amsfonts}
\usepackage{amssymb}

\newcommand{\Ai}{\mathop{\rm Ai}}
\newcommand{\Bi}{\mathop{\rm Bi}}

\font\tenbg=cmmib10 at 10pt
\def \rvecphi{{\hbox{\tenbg\char'036}}}

\begin{document}

\title{Coherent Synchrotron Radiation for Laminar Flows}

\author{Bjoern S. Schmekel}
\affiliation{Theoretical Astrophysics Center, University of California, Berkeley, California 94720}
\email{schmekel@berkeley.edu}

\author{Richard V.E. Lovelace}
\affiliation{Department of Astronomy, Cornell University, Ithaca, New York 14853}
\email{rvl1@cornell.edu}

\begin{abstract}

   We investigate the effect of shear in
the flow of charged particle
equilibria that are unstable to the Coherent
Synchrotron Radiation (CSR) instability.
  Shear may act to quench
this instability because it acts
to limit the size of the region
with  a fixed phase relation
between emitters. 
  The results are important
for the understanding of astrophysical
sources of coherent radiation where shear 
in the flow is likely.

\end{abstract}

\maketitle

\section{Introduction}

When the wavelength of
synchrotron radiation emitted by a
bunch of relativistic particles
is comparable to the size of the bunch
the particles may radiate
coherently. 
  The coherent emission produces
significantly greater power than the
incoherent emission. 
   Coherent synchrotron emission
is seen from 
different astrophysical
source but  most notably from the
rotationally powered radio
pulsars 
(e.g., \cite{ManchesterTaylor1977}).
  In particle
accelerators the coherent 
synchrotron emission is 
usually undesirable because
it  causes very rapid energy loss.
In this context we would also like
to mention efforts to study coherent
curvature radiation as a pulsar
emission mechanism in the laboratory
\cite{benford1981}.

  Several
effects may act to stabilize a system
unstable to the CSR instability. 
 In \cite{Schmekel:2004jb} the authors
analyzed the influence of a small
energy spread in a beam of charged
particles in approximately circular motion.
  A distribution function with a single
value of the canonical angular
momentum was considered. 
   The radial width of
the beam was given by the amplitude
of the betatron oscillations  which
is non-zero for a non-zero energy
spread. 
It was shown that the
decoherence introduced by the
betatron oscillations leads to a
characteristic frequency spectrum,
whereas the dependence on the
Lorentz factor and the number density
remains unaffected.

In this paper we relax the assumption
of zero spread in the canonical
angular momentum $P_{\phi}$ of the
equilibrium distribution. 
   In general 
this leads to shear in that the average
angular velocity becomes dependent on
the radius of the orbit. 
  In practice
the requirement of a small spread in
the canonical angular momentum may be
harder to satisfy than a small
energy spread. 
    Shear is expected
in astrophysical sources of
coherent emission \cite{arons1979}.

Shear itself can be the cause of
instabilities, for example,
the ``diocotron
instability'' \cite{Davidson2001}, but this is not the
focus  of the present paper. 
  In the
case of CSR it is reasonable to
expect that the shear acts to
stabilize the CSR
instability due to 
particles with different
radii ``slipping away''. 
  Using a
linear perturbation analysis of the
fluid equations for a
laminar Brillioun flow we show that
even with a spread of the canonical
angular momentum
$\Delta P_{\phi} > 0$,
previous results for equilibria with
constant $P_\phi$ can be
recovered treating the plasma as a
relativistic cold fluid.
Computer simulations of CSR emission in
Brillioun flows \cite{Schmekel:2004su}  are 
compatible with this picture.
   With the cold
fluid approximation adopted here,
the stability
depends on the number density and the
angular velocity which depends on
the radius.
 The azimuthal velocity is approximately
equal to the speed of light.

The problem of CSR emission has been
investigated by several authors.
Goldreich and Keeley
\cite{GoldreichKeeley1971} considered
the stability of a charge
distribution whose motion is confined
to a thin ring with  the particle motion
being one-dimensional. 
  This calculation led to 
confusion as to how the
proposed CSR mechanism works in detail
(cf. \cite{buschauer1978}).
  
   As proposed in
\cite{Schmekel:2004jb} the CSR
instability is related to the
classical negative mass instability
\cite{Nielson1959,Briggs1966,Entis1971,Kolomenskii1959}
in the sense that an increase in
particle energy leads to a decrease
in its angular velocity. 
   While the classical negative mass
instability is caused by the Coulomb
part of the electromagnetic
potential, the CSR instability is
caused by the radiation field. 
  The negative mass effect is not
immediately apparent in a
one-dimensional treatment based on
conservation of energy and charge.
Heifets and Stupakov
\cite{Heifets:2002un} effectively
built in the negative mass effect by
hand having a constrained radius and
particle energy.

  The treatments
\cite{GoldreichKeeley1971,Heifets:2002un,
Schmekel:2004jb}
give the growth
rate in the absence of an energy
spread. 
  However, considering a
non-zero energy spread requires a
truly two-dimensional model.
Larroche and Pellat investigated the
effect of steep boundaries in the
particle distribution function
\cite{larroche1987}.

Section II describes the assumed  
models, and Section III  approximate
solutions of the equations.
   Section IV derives a dispersion
relation for the considered perturbation.
  Section V discusses the results.  
%
\begin{figure}
\includegraphics[width=3.3in]{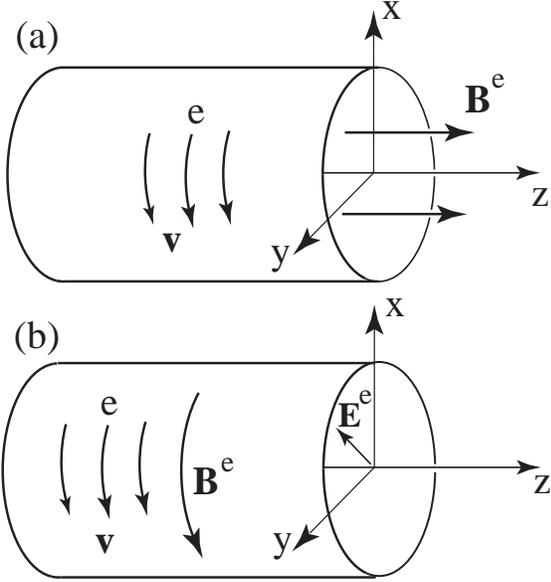}
\caption{Panel (a) shows
the geometry of a relativistic E-layer
for the case of a uniform external 
axial magnetic field.
  Panel (b) shows the second case
of a relativistic E-layer 
in an external toroidal magnetic field
and an external radial electric field.
 }
\label{config}
\end{figure}
%
\section{Theory}
    We consider two cases illustrated in
Figure 1.   The first case (a) corresponds
to a thin layer of relativistic electrons 
gyrating in a uniform external magnetic
field, an ``E-layer.''   The second case (b)
corresponds to the electrons moving almost
parallel to a toroidal magnetic field.

\subsection{Case a}
    We consider a laminar Brillioun
type  equilibrium of a long, non-neutral,
cylindrical relativistic
electron (or positron) layer in a uniform
external magnetic field ${\bf B}_e = B_e\hat{\bf z}$,
where we use a non-rotating cylindrical
$(r,\phi,z)$ coordinate system.
    The electron velocity is
${\bf v} = v_{\phi}(r)\hat{\rvecphi~}=
r\Omega(r)\hat{\rvecphi~}$
   The self-magnetic field is in the
$z-$direction while the self-electric
field is  in the $r-$direction.
   The radial force balance of the equilibrium
is
\begin{equation}
-\gamma \Omega^2 r = {q \over m_e}(E + v B)~,
\label{fbalance}
\end{equation}
where $\gamma = (1-v_\phi^2)^{-1/2}$ is the
Lorentz factor with velocities
measured in units of the speed of light, $B=B_e+B_s$
is the total (self plus external) axial magnetic
field, $E$ is the total ($=$ self) radial
electric field, and $q$ and $m_e$ are the
particle charge and rest mass.
    We have
\begin{equation}
{1\over r}{d (r E) \over dr} = 4 \pi \rho_e~,
\quad
{d B \over dr}= -4\pi \rho_e v~,
\end{equation}
where $\rho_e(r)$ is the charge density of
the electron layer.

    We consider weak layers in the sense that
the `field reversal' parameter
\begin{equation}
\zeta \equiv -~{4\pi \over B_e}
\int_{r_1}^{r_2} dr \rho_e v
\end{equation}
is small compared with unity, $\zeta^2 \ll 1$.
Under this condition equation \ref{fbalance}
gives $\Omega =-qB_e/(m_e \gamma)$.
Here, we have assumed that the layer exists between
$r_1$ and $r_2$.
     We also consider that the Lorentz factor
is appreciably larger than unity in the
sense that $\gamma^2 \gg 1$.
Furthermore, we consider radially thin layers
\begin{eqnarray}
\delta \equiv \frac{\Delta r}{r_0} \ll 1 ~.
\end{eqnarray}

   We consider general electromagnetic
perturbations of the electron layer with
the perturbations proportional
to
\begin{eqnarray}
f_\alpha (r) \exp(im\phi -i\omega t)~,
\label{ansatz}
\end{eqnarray}
where $\alpha =1,2,..$ for the different
scalar quantities, $m=$ integer, and $\omega$
the angular frequency of the perturbation.
   Thus the perturbations  give rise
to field components $\delta E_r$, $\delta E_\phi$,
and $\delta B_z$.
    The perturbed equation of motion is
$$
\left({\partial \over \partial t}
+{\bf v}\cdot{\bf \nabla}\right)
\big({\bf v}\delta \gamma +\gamma \delta {\bf v}\big)
+ \delta {\bf v} \cdot \nabla ( \gamma {\bf v})
$$
\begin{equation}
={q \over m_e} \big(\delta {\bf E}
+ {\bf v} \times \delta {\bf B}
+\delta {\bf v} \times {\bf B} \big)~,
\end{equation}
where the deltas indicate perturbation quantities.
     This equation can be simplified to give
$$
  ~\left [\begin {array}{cc}
{-i\gamma\Delta \omega }&
{-\gamma \Omega(1+\gamma^2) -{q \over m_e} B}
\\\noalign{\medskip}
{\gamma \Omega +(\gamma \Omega r)^\prime +{q\over m_e}B}&
{-i\gamma^3 \Delta \omega}
\end {array}\right ]
\left[\begin{array}{c} {\delta v_r}
\\ \noalign{\medskip}{\delta v_\phi} \end{array}\right]~
$$
\begin{eqnarray}
=~       {q\over m_e}  \left[\begin{array}{c}
{\delta E_r + v \delta B_z}
\\ \noalign{\medskip}{\delta E_\phi} \end{array}\right]~,
\label{simpEuler}
\end{eqnarray}
where the prime denotes a derivative with respect
to $r$, and
$$
\Delta \omega(r) \equiv \omega -m\Omega(r)
$$
is the Doppler shifted frequency seen by a
particle rotating at $\Omega$.
We also define the dimensionless quantity
\begin{eqnarray}
\Delta \tilde \omega \equiv \frac{\Delta \omega}{m \Omega}
\end{eqnarray}
which will turn out to be useful later.

    Using the equilibrium equation \ref{fbalance} and
the condition $\zeta^2 \ll 1$, the matrix in
equation \ref{simpEuler} is approximately
\begin{equation}
  {\cal D}= \left [\begin {array}{cc}
{-i\gamma\Delta \omega }&
{-\gamma^3 \Omega }
\\\noalign{\medskip}
{\gamma^3(\Omega r)^\prime }&
{-i\gamma^3 \Delta \omega}
\end {array}\right ]~,
\end{equation}
    We have used the fact that
$(\gamma\Omega r)^\prime =
\gamma^3(\Omega r)^\prime$.
For $\zeta^2 \ll 1$ we have $(\Omega r)^\prime
=\Omega /\gamma^2$ and $\Omega^\prime = -v^2\Omega/r$.
In the absence of shear the latter quantity would be zero.
    Consequently
\begin{equation}
{\rm det}({\cal D}) =
\gamma^4(\Omega^2 -\Delta \omega^2)~.
\end{equation}
Inverting equation \ref{simpEuler} gives
\begin{equation}
\delta v_r = {q \gamma^3 \over m_e {\rm det}({\cal D})}
\bigg[-i\Delta \omega (\delta E_r +v \delta B_z)
+ \Omega \delta E_\phi \bigg]~,
\label{dvr}
\end{equation}
and
\begin{equation}
\delta v_\phi = {q \gamma\over m_e {\rm det}({\cal D})}
\bigg[-i\Delta \omega \delta E_\phi -
  \Omega (\delta E_r + v\delta B_z) \bigg]~.
\label{dvphi}
\end{equation}
Here, $q \rho_e \gamma^3 \Omega / (m_e {\rm det}({\cal D}))$
has the role of the distribution function of angular momentum
\cite{lovelace1979}.
\subsection{Case b} 
Here we consider
an equilibrium with the same
number density and velocity profile 
as before but with different external fields
as shown in Figure \ref{config}b.
  Instead
of an external magnetic field in the 
$z-$direction we consider an equilibrium
with an azimuthal magnetic field acting as a guiding field and 
a radial electric field. The latter is included in the equilibrium
condition and therefore does not 
enter the linearized Euler equation. $B^e_\phi$ would only enter
if we considered motion in the 
axial direction and non-zero axial wavenumbers. 

Thus, we obtain
the matrix $\mathcal{D}$ again 
without the $B_0$ terms, i.e. for $\gamma \gg 1$
\begin{equation}
  {\cal D}= \left [\begin {array}{cc}
{-i\gamma\Delta \omega } & {-\gamma^3 \Omega }
\\\noalign{\medskip}
{2 \gamma \Omega} & {-i\gamma^3 \Delta \omega}
\end {array}\right ]~,
\end{equation}
with
\begin{equation}
{\rm det}({\cal D}) =
\gamma^4(2 \Omega^2 -\Delta \omega^2)~.
\end{equation}
We obtain
\begin{equation}
\delta v_r = {q \gamma^3 \over m_e {\rm det}({\cal D})}
\bigg[-i\Delta \omega (\delta E_r +v \delta B_z)
+ \Omega \delta E_\phi \bigg]~,
\label{dvr2}
\end{equation}
and
\begin{equation}
\delta v_\phi = {q \gamma\over m_e {\rm det}({\cal D})}
\bigg[-i\Delta \omega \delta E_\phi -
  2 \Omega (\delta E_r + v\delta B_z) \bigg]~.
\label{dvphi2}
\end{equation}
Such a configuration is a more
realistic possibility for the 
magnetosphere of a radio pulsar.
The conclusions for the two configurations do not differ significantly,
and we will proceed analyzing case a.

\section{Approximate Solution}
Using equations (\ref{dvr}) and (\ref{dvphi})
the linearized continuity equation gives
\begin{eqnarray} \nonumber
& i &  \!\!\! \Delta \omega \delta \rho = \hspace{2.5in}
\\ \nonumber
& + & \!\!\! \frac{im}{r} \left [ \frac{q \gamma \rho_0}{m_e {\rm det}({\cal D})} \left (-i \Delta \omega \delta E_\phi - \Omega
\left (\delta E_r + v_\phi \delta B_z \right ) \right ) \right ]
\\  \nonumber
& + & \!\!\! \frac{1}{r} \frac{\partial}{\partial r} \left [ r \frac{q \gamma^3 \rho_0}{m_e {\rm det}({\cal D})} \left (-i \Delta \omega  \left (\delta E_r + v_\phi \delta B_z \right ) + \Omega
\delta E_\phi \right ) \right ]  ~,
\\
\end{eqnarray}
where we used $\delta J_r = \rho_0 \delta v_r$ and $\delta J_\phi = v_\phi \delta \rho + \rho_0 \delta v_\phi$.
We consider conditions where $\delta E_r + v_\phi \delta B_z$ terms can be neglected.
The sufficient conditions are
\begin{eqnarray}
\left | \frac{\Delta \omega}{\Omega} \right | \le 1 ~,
\label{DomOmega}
\end{eqnarray}
and
\begin{eqnarray}
\left | \delta E_r + v \delta B_z \right | \ll \left | \frac{\Delta \omega}{\Omega} \right |
\left | \delta E_{\phi} \right | ~.
\label{smallvphi}
\end{eqnarray}
We estimate the relative magnitude of the field components using the Maxwell equations.
From Faraday's law we obtain the relation
\begin{eqnarray}
D_r (\delta E_{\phi}) - i k_{\phi} \delta E_r = i \omega \delta B_z
\end{eqnarray}
where $D_r (...) \equiv r^{-1} \partial / \partial r (r ...) \approx i k_r$
assuming the radial dependence $\exp (i k_r r)$ for the perturbed quantities
as well as thin layers with $\delta \ll 1$. We have
\begin{eqnarray}
k_r \delta E_{\phi} = k_{\phi} \left ( \delta E_r + v_{\phi} \delta B_z \right ) + k_{\phi} v_{\phi} \Delta \tilde \omega \delta B_z
\end{eqnarray}
With $\delta B_z = D_r \delta A_{\phi}$, $\delta A_{\phi} = v_{\phi} \delta \Phi$ and $\delta E_{\phi}=-imr^{-1} \delta \Phi + i \omega \delta A_{\phi}$ 
\begin{eqnarray}
\delta E_r + v_{\phi} \delta B_z = \frac{k_r}{k_{\phi}} \delta E_{\phi} - v_{\phi}^3 \Delta \tilde \omega k_r \frac{\delta E_{\phi}}{\Delta \omega - 
\omega / \gamma^2} \approx \frac{\bar k_r}{m} \delta E_{\phi}
\end{eqnarray}
Inequality (\ref{smallvphi}) turns into
\begin{eqnarray}
\bar k_r \ll m \frac{\Delta \omega}{\omega} = m^2 \Delta \tilde \omega
\end{eqnarray}
where $\bar k_r \equiv k_r r_0$.
We will always assume radial wavenumbers $k_r$ that are sufficiently small
such that the latter condition is met.
A consequence of inequality (\ref{smallvphi}) is that
\begin{eqnarray}
\delta v_{\phi} = \gamma^{-2} \frac{\Delta \omega}{\Omega} \delta v_r
\end{eqnarray}

Making use of approximation \ref{smallvphi}, equation~\ref{dvr} 
can be entirely written in terms of $\delta E_{\phi}$.
Note that the resonant term due to $\det({\mathcal{D})}$ 
(the Lindblad resonance) is canceled in the limit
$\delta v_{\phi} \longrightarrow 0$. In general we will have
to invoke the assumptions (\ref{DomOmega}) and (\ref{smallvphi}), though. 
Neglecting the small $\delta v_{\phi}$ term,  the 
linearized continuity equation gives
\begin{eqnarray}
\frac{1}{r} \frac{\partial}{\partial r} 
\left(r \delta J_r\right) 
= i \Delta \omega \delta \rho~.
\end{eqnarray}
Thus,
\begin{eqnarray}
r \delta \rho = \frac{-i}{\Delta \omega} 
\frac{\partial}{\partial r} \left(
\frac{q r \rho_0 }{m_e \gamma \Omega} 
\delta E_{\phi} \right)~.
\label{intrdrho}
\end{eqnarray}
Integrating over $dr$ and integrating by parts gives
\begin{eqnarray}
\int rdr \delta \rho  = 
- \int dr \frac{im \Omega/r}{(\Delta \omega)^2}
\frac{q r \rho_0}{m_e \gamma \Omega} \delta E_{\phi}
\end{eqnarray}

In the Lorentz gauge
\begin{eqnarray}
\delta E_{\phi} &= &- \frac{im}{r} 
\delta \Phi + i \omega \delta A_{\phi}
\nonumber\\
\delta E_r &=& - \frac{\partial}{\partial r} 
\delta \Phi + i \omega \delta A_r
\nonumber\\
\delta B_z &=& \frac{1}{r} \frac{\partial}
{\partial r} \left( r \delta A_{\phi} \right)~. 
\end{eqnarray}
For $(\Delta r/r)^2 \ll 1$ we have
the approximation 
$r^{-1}\partial[r(\partial.../\partial r)]
=\partial(...)/\partial r$.
Assuming the radial 
dependence $\delta \Phi \propto
\exp (i k_r r)$ we obtain
from the gauge condition 
\begin{eqnarray}
i k_r \delta A_r + \frac{im}{r} \delta A_{\phi} - i \omega \delta \Phi = 0
\end{eqnarray}
and $\delta A_z$ is negligible because of the symmetry of the problem. $\delta A_r$ is non-zero
since $\delta J_r =  \rho_0 \delta v_r + v_r \delta \rho \neq 0$ and 
$|\delta v_{\phi}| \sim \gamma^{-2} |\Delta \omega / \Omega| |\delta v_r|$.
$\delta A_{\phi}$ can be computed from the Green's
function (cf. appendix) to give $\delta A_{\phi} = v_{\phi} \delta \Phi$.
Fortunately, we only need $\delta E_{\phi}$.
Note that
\begin{eqnarray}
\delta E_{\phi} = -im r^{-1} \delta \Phi 
/ \gamma^2 + O(\Delta  \tilde \omega)
\end{eqnarray}

\section{Dispersion relation}
The derivation of the Green's function can be found in the appendix
\begin{eqnarray}
\delta \Phi(r) = 2 \pi^2 i 
\int_0^{\infty} r' dr' J_m(\omega r_{<}) 
H_m (\omega r_{>}) \delta \rho(r')
\end{eqnarray}

The argument of the Bessel functions is assumed to be
independent of $r$ and $r'$ in the important region
$1-\delta < r/r_0 < 1 + \delta$ which we will justify later.
Thus, 

\begin{eqnarray}
\delta \Phi(r) = - \frac{2 \pi^2 i e^2 m^2}
{H \gamma^2} J_m H_m \int_0^{\infty} dr' 
\frac{n(r') \delta \Phi(r')}{r' (\omega
- m \Omega(r'))^2}
\end{eqnarray}
where Eq.~\ref{intrdrho} has been used.
Since the right hand side of the last 
equation is independent of $r$ $\Phi(r)$ 
has to be constant and we obtain with
$\Omega(r) \approx 1/r$
\begin{eqnarray}
1 = - \frac{2 \pi^2 i e^2 m^2}{H \gamma^2} 
J_m H_m \int_0^{\infty} dr' \frac{n(r')}{r' (\omega-m/r')^2}
\end{eqnarray}
where the Bessel functions are evaluated 
at $m(1-1/(2\gamma^2))$. The remaining integral can be
evaluated if the Gaussian number density profile
$n(r) = n_0 \exp \left [ - \delta r^2/2 (\Delta r)^2 \right ]$
is replaced by a rectangle with 
width $2 \Delta r$ and height
$n_0 \sqrt{\pi/2}$. 
\begin{eqnarray} \nonumber
\int_0^{\infty} dr' \frac{n(r')}{r' (\omega - m/r')^2} \approx \int_{r_0 (1-\delta)}^{r_0 (1+\delta)} dr' n_0 \frac{\sqrt{\pi/2}}{r' (\omega - m/r')^2} \quad
\\ \nonumber
= n_0 \sqrt{\frac{\pi}{2}} \left [ \frac{\ln(\omega r' - m)}{\omega^2} - \frac{m}{\omega^2 (\omega r' - m)} \right ]_{r_0(1-\delta)}^{r_0(1+\delta)}
\\
\approx n_0 \sqrt{\frac{\pi}{2}} \frac{2 r_0 \delta}{\omega} \frac{m}{(\omega r_0 - m)^2-\omega^2 r_0^2 \delta^2} \quad \quad
\label{evalint}
\end{eqnarray}
The logarithm can be neglected because we are interested in the resonant case.
With $\zeta = 4\pi e^2 n_0 \sqrt{\pi/2} r_0^2 \delta/m_e \gamma$ we obtain the dispersion relation
\begin{eqnarray}
1 = - \pi \zeta Z \gamma^{-2} \frac{1}{(\Delta \tilde \omega)^2 - \delta^2} 
\end{eqnarray}
where $Z = i J_m H_m$. Thus,
\begin{eqnarray}
\Delta \tilde \omega = 
\pm \sqrt{\delta^2 - \pi \zeta Z / \gamma^2}
\end{eqnarray}
The Bessel functions can be expressed in terms of Airy functions for $m \gg 1$
\begin{eqnarray}
J_m(z) = & 2^{1/3} & m^{-1/3} \Ai \left ( w \right )
\label{JmAi}
\end{eqnarray}
\begin{eqnarray}
Y_m(z) = - & 2^{1/3} & m^{-1/3} \Bi \left ( w \right )
\label{YmBi}
\end{eqnarray}
where
\begin{eqnarray} \nonumber
w & \equiv & - 2^{1/3} m^{2/3} \left ( \frac{z}{m} - 1 \right ) \\
    & = & \left ( \frac{m}{2} \right )^{2/3} \left [ \gamma^{-2} - 2 \Delta \tilde \omega - 2 (\xi - 1) \right ]
\end{eqnarray}
with $\xi \equiv r/r_0$ which at the outer edge is $\xi - 1 \sim \delta$.
Assuming $m \ll 2 \gamma^3 \equiv m_2$, $\gamma^{-2} \ll 1$, $|\Delta \tilde \omega| \ll \gamma^{-2}$
(assuming that the layer essentially exists in the region $1-\delta < r/r_0 < 1 + \delta$.) and 
$2^{1/3} \delta \ll m^{-2/3}$
we can consider $Z$ to be a pure function of $m$ since $|w|^2 \ll 1$. 
\begin{eqnarray}
Z \approx (0.347 + 0.200i)/m^{2/3}
\end{eqnarray}
Therefore, the Bessel functions do not depend on $r$ and $r'$ anymore.
For $|w| \gg 1$ the Airy functions can be approximated to give
\begin{eqnarray}
Z \approx \frac{1}{2 \pi \sqrt{w}} \left ( \frac{2}{m} \right )^{2/3} ~,
\end{eqnarray}
but we additionally have to demand $\delta \ll \gamma^{-2}$ in order for $Z$ to remain independent of $r$ and $r'$.
For zero energy spread the condition $|\Delta \tilde \omega| \ll \gamma^{-2}$ is equivalent to $m \ll \gamma^3 \zeta^{3/2} \equiv m_1$.
As we will discuss in Section V this condition may be too strict.

\begin{figure}
\includegraphics[width=3.5in]{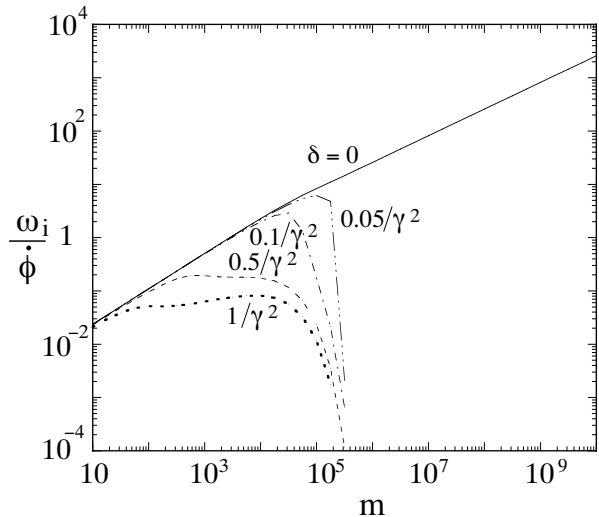}
\caption{Growth rates as a function 
of azimuthal mode $m$ for $\zeta=0.02$, $\gamma=30$
and various energy spreads.}
\label{thick}
\end{figure}

Some results are plotted in Fig.~\ref{thick}. For zero energy spread we recover the
usual scaling relation $\Im(\omega)/\Omega(r_0) \propto m^{2/3}$ for $m<2 \gamma^3$
\cite{Schmekel:2004jb, GoldreichKeeley1971, Heifets:2002un} and $\Im(\omega)/\Omega(r_0) \propto m^{1/3}$
for $m>2 \gamma^3$ \cite{Schmekel:2004jb} with the $|w| \gg 1$ approximation of the Airy functions. The most dramatic consequence of a non-zero
energy spread is the presence of a very sudden and steep cut-off. Pushing the cut-off
to higher values of $m$ requires increasingly small energy spreads. 
For large energy spreads the growth rate $\Im(\omega)/\Omega(r_0)$ scales
as $m^{1/3}$ if the $|w| \ll 1$ approximation is used. Retaining the Airy function the growth rate
grows more slowly. According to Fig.~\ref{thick} the scaling relation for $\delta=1/\gamma^2$
is in the order of $m^{1/4}$ before the cut-off is reached.
This power law has been derived before in
\cite{Schmekel:2004jb} where the decoherence is due to betatron oscillations
instead of a non-zero spread in the average angular velocity. Note that these two
results also agree with computer simulations that were carried out for Brillioun 
\cite {Schmekel:2004su} flows.
%
%
%
\section{Results and Discussion} 
In order for the growth rate to vanish 
the expression inside the root must be real
and non-negative. Using our approximations 
for the Bessel functions it is seen immediately
that the former condition is satisfied 
if $m>m_2$. This explains why the drop off
occurs at $m=m_2$ regardless 
of the value of $\delta$ as long as the remaining real
part is positive. 
This is the case if the field reversal parameter does not
exceed the critical threshold
\begin{eqnarray}
\zeta > m \gamma \delta^2  ~.
\label{zetathreshold}
\end{eqnarray}
If the last inequality is not satisfied complex roots can only exist if $Z$ is complex  which is the case for $m < m_2$.
If however this inequality is satisfied the expression inside the root becomes negative
for $m > m_2$ and an unstable solution exists, but the sharp cut-off is replaced by Eq.~\ref{zetathreshold}.

As discussed in the last section the azimuthal 
mode number $m$ resulting in the largest growth rate
is either given by $m=2 \gamma^3$ or $m=\zeta \gamma^{-1} \delta^{-2}$ whatever is greater.
This guarantees that $(\Delta \tilde \omega)^2$ is either still complex or real and negative with both
leading to an unstable mode. For our allowed parameter range $2 \gamma^3$ is typically larger.
Thus,
\begin{eqnarray}
\Im(\omega) \le 2^{2/3} \dot \phi \gamma \sqrt{\zeta} \sim \gamma^{-1/2} \omega_p
\end{eqnarray}
 where $\omega_p$ is the nonrelativistic plasma frequency.

\begin{figure}
\includegraphics[width=3.3in]{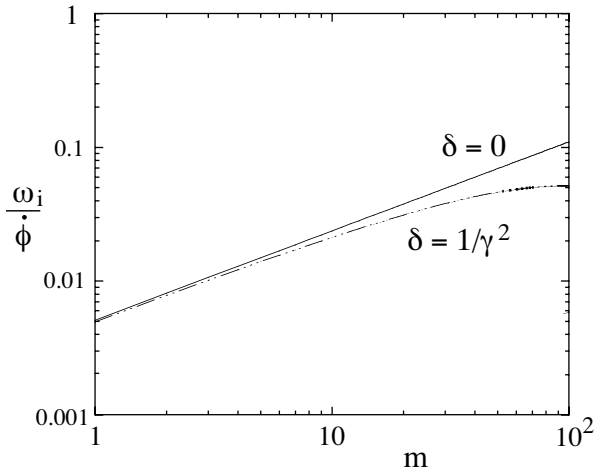}
\caption{Growth rates as a function of 
azimuthal mode $m$ for $\zeta=0.02$, $\gamma=30$
and various energy spreads. 
The Bessel functions were retained in order to compute
the growth rates in the low $m$ regime correctly.}
\label{smallm}
\end{figure}

   Finally, let us consider small values
of the azimuthal mode number $m$.
  So far we have have
approximated the  Bessel functions by Airy functions
which makes them easier to 
compute especially for large orders. For small
$m$ this approximation cannot 
be justified and the Bessel functions
have to be retained. 
In Figure \ref{smallm} we 
solved the dispersion relation
in the small $m$ regime. 
Also, their arguments depend on $\Delta \tilde \omega$ since $m<m_1$.
An accurate calculation 
of the growth rates of
modes with azimuthal mode nubers 
$m$ is important for the determination 
of the total power radiated, because 
the latter decreases with increasing $m$. 
Even for $m=1$ Eq.~\ref{JmAi} and \ref{YmBi} give excellent results, presumably
because the argument of the Bessel functions is close to their order for $\gamma \gg 1$
and $Z$ is independent of $w$ if $|w|^2 = \left | \left ( \frac{m}{2} \right )^{2/3}
\left [ \gamma^{-2} - 2 \Delta \tilde \omega - 2 (\xi-1) \right ] \right |^2 \ll 1$.
The latter condition may be satisfied even if $|\Delta \tilde \omega|$ violates our
assumption $|\Delta \tilde \omega| \ll \gamma^{-2}$.

Looking at Eq.~\ref{evalint} we conclude that the quenching of the instability, i.e.
the existence of the additional $\delta^2$ term in the denominator, is due to
both the non-zero thickness of the layer and $\Omega' \neq 0$. However, it is
important to note that the instability relies on the negative mass effect and
would not exist in the absence of shear in our model.

\acknowledgments
We thank Ira M.~Wasserman and Georg H.~Hoffstaetter for many valuable discussions.
This research was supported by the Stewardship 
Sciences Academic Alliances program of the National Nuclear
Security Administration under US Department of Energy Cooperative agreement DE-FC03-02NA00057.

\appendix*
\section{Green's Function}
The Green's function for the potentials give
\begin{eqnarray}
\delta \Phi({\bf r},t)=
\int dt^\prime d^3 r^\prime ~
G({\bf r}-{\bf r^\prime},t-t^\prime) ~
\delta \rho({\bf r}^\prime,t^\prime)~.
\nonumber\\
\delta {\bf A}({\bf r},t)=
\int dt^\prime d^3 r^\prime ~
G({\bf r}-{\bf r^\prime},t-t^\prime) ~
\delta {\bf J}({\bf r}^\prime,t^\prime)~,
\end{eqnarray}
where
$$
\left(\nabla^2 - {\partial^2 \over \partial t^2}\right)
G({\bf r},t)= -4\pi \delta(t)\delta({\bf r})~,
~~ \tilde{G}({\bf k},\omega)=
{4\pi \over  {\bf k}^2-\omega^2  }~,
$$
\begin{eqnarray}
G({\bf r},t)=4\pi\int_C d\omega \int d^3k~
{\exp(i{\bf k\cdot r}-i\omega t) \over  {\bf k}^2-\omega^2 }~,
\end{eqnarray}
where $\tilde{G}$ is the Fourier transform of the Green's
function. The ``C'' on the integral indicates an
$\omega-$integration parallel to but above the real axis, ${\rm
Im}(\omega)>0$, so as to give the retarded Green's function.

        Because of the assumed dependences
of Eq. (\ref{ansatz}),
we  have for the electric potential,
$$
\delta \Phi_{\omega m k_z}(r) = 2
\int_0^\infty r^\prime dr^\prime
\int_0^\infty \kappa d\kappa \int_0^{2\pi} d\alpha~
~ \delta
\rho_{\omega m k_z}(r^\prime)\big[.. \big]
$$
\begin{equation}
=4\pi \int_0^\infty r^\prime dr^\prime
\int_0^\infty \kappa d\kappa~
{J_m(\kappa r) J_m(\kappa r^\prime)
\over \kappa^2- (\omega^2 - k_z^2) }~
\delta\rho_{\omega m k_z}(r^\prime)~,
\end{equation}
where
$$
\big[.. \big]\equiv{\exp(im\alpha)
J_0\{\kappa [r^2+ (r^\prime)^2-2r
r^\prime
\cos\alpha]^{1/2}\}
\over \kappa^2-(\omega^2 - k_z^2) }~,
$$
where $\kappa^2 \equiv k_x^2+k_y^2$.
Because $\omega$ has a positive imaginary part,
this solution corresponds to the retarded field.
      Also because ${\rm Im}(\omega)>0$, the
$\kappa-$integration can be done by
a contour integration as discussed in \cite{Watson1966}
which gives
\begin{equation}
\delta \Phi_{\omega m k_z}(r)\!=\!
2\pi^2 i \int_0^\infty \!\!\!r^\prime dr^\prime
J_m(kr_<)H_m^{(1)}(kr_>)\delta\rho_{\omega m k_z}(r^\prime),
\label{dpot1}
\end{equation}
where $k\equiv  (\omega^2-k_z^2)^{1/2}$, where $r_<$ ($r_>$)
is the lesser (greater) of $(r,r^\prime)$, and where
$H_m^{(1)}(x)=J_m(x)+iY_m(x)$ is the  Hankel
function of the first kind.
Because $(\delta A_\phi,\delta J_\phi)=
-(\delta A_x,\delta J_x)\cos\phi+(\delta A_y,
\delta J_y)\sin\phi$, we have instead of Eq. (\ref{dpot1}),
\begin{eqnarray}
\nonumber
\delta \Psi^{\omega m k_z}(r)= r\delta A_\phi^{\omega m k_z}=
\pi^2 r i \int_0^\infty\!\!\! r^\prime dr^\prime
\delta
J_\phi^{\omega m k_z}(r^\prime)
\\ \cdot
J_{m+1}(kr_<) H_{m+1}^{(1)}(kr_>)+
J_{m-1}(kr_<) H_{m-1}^{(1)}(kr_>).
\label{dpot2}
\end{eqnarray}
For $m \gg 1$, one can show that in this equation
$[J_{m+1}H_{m+1}^{(1)}+
J_{m-1}H_{m-1}^{(1)}]/2 = J_m H_m^{(1)}$
to a good approximation.
Equations (\ref{dpot1}) and (\ref{dpot2}) are useful in subsequent
calculations.

\end{document}